\documentclass[twocolumn,prl,a4paper,showpacs,floatfix,preprintnumbers,amsmath,amssymb]{revtex4}

\usepackage{amsmath}
\usepackage{graphicx}

\usepackage{graphicx}
\usepackage{dcolumn}
\usepackage{bm}

\newcommand{\ud}{\mathrm{d}}
\newcommand{\be}{\begin{equation}}
\newcommand{\ee}{\end{equation}}

\begin{document}

\title{Phase detection at the quantum limit with multi-photon Mach-Zehnder interferometry}

\pacs{42.50 St, 42.50 -p}

\author{L. Pezz\'e, A. Smerzi}
\affiliation{ BEC-CNR-INFM and Dipartimento di Fisica, Universit\`a di Trento, I-38050 Povo, Italy}
\author{G. Khoury, J. F. Hodelin and D. Bouwmeester}
\affiliation{ Department of Physics, University of California, Santa Barbara,
California 93106, USA}



\begin{abstract}
We study a Mach-Zehnder interferometer fed by a coherent state in
one input port and vacuum in the other. We explore a Bayesian
phase estimation strategy to demonstrate that it is possible to
achieve the standard quantum limit independently from the true
value of the phase shift and specific assumptions on the noise of
the interferometer.  We have been able to implement the protocol
using parallel operation of two photon-number-resolving detectors
and multiphoton coincidence logic electronics at the output ports
of a weakly-illuminated Mach-Zehnder interferometer. This protocol
is unbiased and saturates the Cramer-Rao phase uncertainty bound
and, therefore, is an optimal phase estimation strategy.  
\end{abstract}

\maketitle

\begin{figure}[t!]
\begin{center}
\includegraphics[scale=0.55]{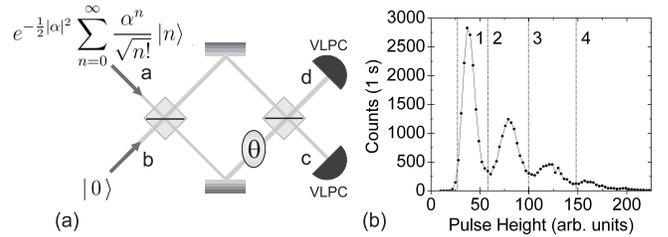}
\end{center}
\caption{\small{(a) Schematic of a Mach-Zehnder interferometer. A
phase sensitive measurement is provided by the detection of the
number of particles $N_c$ and $N_d$ at the two output ports. (b)
Pulse height distribution for a visible light photon counter
(VLPC) used in the experiment. The power incident on the detector
is $144$ fW at a wavelength of 780 nm. The vertical lines show
the decision thresholds \cite{Khoury_2006}.}}\label{MZ}
\end{figure}

The Mach-Zehnder (MZ) interferometer~\cite{Mach,Zehnder} is a
truly ubiquitous device that has been implemented using photons,
electrons~\cite{Ji_2003}, and atoms~\cite{berman,mm}. Its
applications range from micro- to macro-scales, including models
of aerodynamics structures, near-field scanning microscopy
\cite{zenhausern_1995} and the measurement of gravity
accelerations \cite{peters_1999}. The central goal of
interferometry is to estimate phases with the highest possible
confidence \cite{Holland_1993, Sanders_1995, Pezze_2006} while
taking into account sources of noise. Recent technological
advances make it possible to reduce or compensate the classical
noise to the level where a different and irreducible source of
uncertainty becomes dominant: the quantum noise. Given a finite
energy resource, quantum uncertainty principles and back reactions
limit the ultimate precision of a phase measurement. In the
standard configuration of the MZ interferometer, a coherent
optical state with an average number of photons $\bar{n}=|\alpha|^2$ enters
input port $a$ and the vacuum enters input port $b$, as
illustrated in Figure~(\ref{MZ}). The goal is to estimate the
value of the phase shift $\theta$ after measuring a certain number
of photons $N_c$ and $N_d$ at output ports $c$ and $d$, which, in
the experiment discussed in this Letter, is made possible by two
number-resolving photodetectors.

The conventional phase inference protocol estimates the true value
of the phase shift $\theta$ as \cite{Sculli, Dowling_1998,
Caves_1981}: \be \label{estimator} \Theta_\text{est} =
\arccos\Big(\frac{M_p}{\bar{n}}\Big), \ee where
\mbox{$M_p=\sum_{k=1}^{p}(N_c^{(k)}-N_d^{(k)})/p$} is the photon
number difference detected at the output ports, averaged over $p$
independent measurements. The phase uncertainty of estimator
(\ref{estimator}) is \be \label{momentum_formula} \Delta \Theta =
\frac {1} {\sqrt{p\, \bar{n}}\sin \theta}, \ee which follows from
a linear error propagation theory. Eq.(\ref{momentum_formula})
predicts an optimal working point at a phase shift $\theta =
\pi/2$, where the average photon number difference varies most
quickly with phase. As $\theta$ approaches 0 or $\pi$, the
confidence of the measurement becomes very low and eventually
vanishes.
As a consequence, this interferometric protocol does not allow the
measurement of arbitrary phase shifts. This can be a serious
drawback for applications like laser gyroscopes, the
synchronization of clocks, or the alignment of reference frames.
Furthermore, to estimate small phase shifts with the highest
resolution, the interferometer has to be actively stabilized around
$\pi/2$.
This generally requires the addition of a feedback loop, which can be quite
costly in terms of time and energy resources.

\begin{figure}
\begin{center}
\includegraphics[scale=0.5]{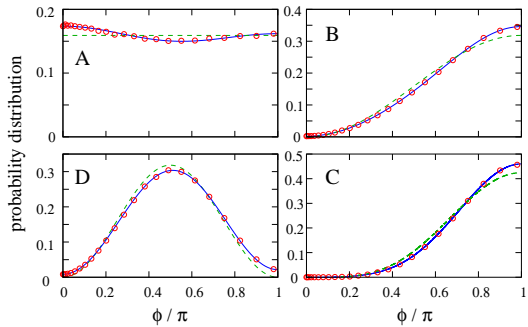}
\end{center}
\caption{\small{(color online). Phase distribution $P(\phi|N_c,
N_d)$, for: A) $N_c=0$, $N_d=0$; B) $N_c=0$, $N_d=1$;  C) $N_c=1$,
$N_d=1$; D) $N_c=0$, $N_d=2$. The circles are the experimental
data collected in the calibration part of the experiment, the
dashed line is the ideal phase distribution
Eq.(\ref{coherent_dist}). The solid line is a fit of the data
according to Eq.(\ref{fitP}).}}\label{R}
\end{figure}
It was first noticed by Yurke, McCall and Klauder (YMK) in
\cite{Yurke_1986} that the estimator (\ref{estimator}) does not
take into account all the available information, and, in
particular, the fluctuations in the total number of photons at the
output ports. The possibility to improve
Eq.(\ref{momentum_formula}) is confirmed by the analysis of the
Cramer-Rao lower bound (CRLB) \cite{Cramer, Rao}, which provides,
given an input state and choice of observables, the lowest
uncertainty allowed by Quantum Mechanics. For a generic, unbiased,
estimator, \mbox{$\Delta \Theta_{CRLB} = 1/\sqrt{p F(\theta)}$},
where $F(\theta)$ is the Fisher information \cite{Fisher}, which,
in general, can depend on the true value of the phase shift
$\theta$ and the number of independent measurements $p$. A direct
calculation of the Fisher information for the coherent $\otimes$
vacuum input state, gives $F(\theta)= \bar{n}$. Therefore, the
Cramer-Rao lower bound is
\be \label{shot_noise}
\Delta \Theta_{CRLB} = \frac{1}{\sqrt{p \, \bar{n}}},
\ee
which, in
contrast with the result of Eq.(\ref{momentum_formula}), is
independent of the true value of the phase shift. The only
assumption here is that the observable measured at the output
ports is the number of particles. It is well known (see for
instance \cite{Helstrom}) that the Maximum Likelihood (ML)
estimator, defined as the maximum, $\Theta_{ML}$, of the
Likelihood function $P(N_c,N_d|\phi)$ (see below), saturates the
CRLB, but only asymptotically in the number of measurements $p$.
In the current literature there have been alternative suggestions
to obtain an unbiased estimator and a phase independent
sensitivity with a Mach-Zehnder interferometer \cite{Yurke_1986,
Hradil_1996, Berry_2000, Sanders_1995, NFM_1991}. They will be specifically
addressed at the end of this Letter.

Here we develop a protocol based on a Bayesian analysis of the
measurement results
\cite{Helstrom,Holland_1993,Hradil_1996,Pezze_2006}. The goal is
to determine $P(\phi | \, N_c, N_d)$, the probability that the
phase equals $\phi$ given the measured $N_c$ and $N_d$. Bayes'
theorem provides this: \mbox{$P(\phi| N_c, N_d) = P(N_c, N_d|\phi)
~P(\phi)/  P(N_c, N_d)$}, where $P(N_c, N_d| \phi)$ is the
probability to detect $N_c$ and $N_d$ when the phase is $\phi$
\cite{nota2}, $P(\phi)$ quantifies our prior knowledge about the
true value of the phase shift, and $P(N_c, N_d)$ is fixed by
normalization. Assuming no prior knowledge of the phase shift,
$P(\phi)=1/\pi$. In the ideal case, the Bayesian phase probability
distribution can be calculated analytically for any value of $N_c$
and $N_d$, \be \label{coherent_dist} P(\phi|N_c, N_d) =
\mathcal{C} \Big( \cos \frac{\phi}{2} \Big)^{2N_c} \Big( \sin
\frac{\phi}{2} \Big)^{2N_d}, \ee where
$\mathcal{C}=\frac{\Gamma(1/2+N_c)\Gamma(1/2+N_d)}{\Gamma(1+N_c+Nd)}$
is a normalization constant. In practice, one must measure $P(N_c,
N_d|\phi)$ and, from this, determine $P(\phi | N_c, N_d)$. This
distribution provides both an estimate on the phase and the
uncertainty in this estimate.

\begin{figure}
\begin{center}
\includegraphics[scale=0.5]{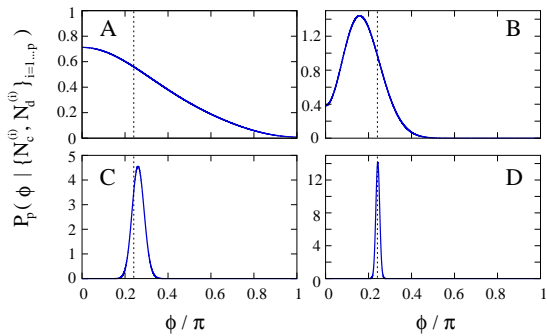}
\end{center}
\caption{\small{(color online). Phase probability distribution Eq.(\ref{Pp}),
obtained after $p$ independent experimental measurements
$\{N_c^{(i)}, N_d^{(i)} \}_{i=1..p}$: A) $p=1$, B) $p=10$, C)
$p=100$, D) $p=1000$. The true value of the phase shift is
$\theta/\pi=0.24$, shown by the vertical dashed line.}}\label{p}
\end{figure}

There are several advantages to using a Bayesian protocol.
Notably, it can be applied to any number $p$ of independent
measurements, it does not require statistical convergence or
averaging, and it provides uncertainty estimates tailored to the
specific measurement results. For instance, with a single
measurement, $p=1$, it predicts an uncertainty that scales as
$\Delta \Theta \approx \frac{1}{\sqrt{N_c + N_d}}$.
Since Eq.(\ref{coherent_dist}) does not depend on $\bar n$, the
estimation is insensitive to fluctuations of the input laser
intensity. Most importantly, its uncertainty, in the limit $p>> 1$, is
$\Delta \Theta =\frac{1}{\sqrt{p\,\bar{n}}}$ which
coincides with the CRLB. 

To implement the proposed protocol we have realized a polarization
Mach-Zehnder interferometer with photon-number-resolving
coincidence detection.
In a recent paper we reported on the analysis of a coherent state
using a single photon-number-resolving detector 
\cite{Khoury_2006}. We have extended this experimental capability
to two simultaneously operating visible light photon counters
(VLPCs)~\cite{Turner}, cryogenic photodetectors that provide a
current pulse of approximately 40,000 electrons per detected
photon. The VLPCs were maintained at 8~K in a helium flow
cryostat, and their photocurrent was amplified by low-noise, room
temperature amplifiers. We measured a detection efficiency of 35\%
and a dark count rate of $3\times10^5$ for each detector under our
operating conditions. Custom electronics processed the amplified
VLPC current pulses to perform gated, fast coincidence detection.
We were thus able to determine, for each pulse, how many photons
were detected at both ports $c$ and $d$. A Ti:sapphire
pulsed laser, attenuated such that $\bar{n} =
1.08$ photons, provided the input state. Since a coherent state
maintains its form under linear loss~\cite{Sculli}, the presence
of loss after the interferometer is completely equivalent to a
lossless interferometer fed by a weaker input state. We use
$\bar{n}$ to signify the average number of photons in the detected
state per pulse, after all losses. We were limited by the
amplifiers to measuring up to four photons per pulse
Fig.~\ref{MZ}(b), but at $\bar{n} = 1.08$ the probability of
detecting five or more photons is negligible. The phase shift
$\theta$ was changed by tilting a birefringent crystal inside the
interferometer.

\begin{figure}
\begin{center}
\includegraphics[scale=0.34]{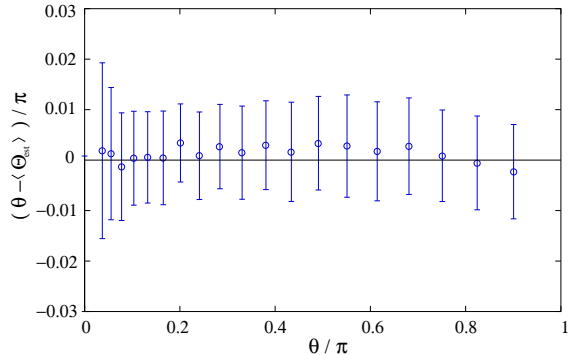}
\end{center}
\caption{\small{(color online). Difference between the mean value
of the phase estimator, $\langle \Theta_\text{est} \rangle$,
obtained after 150 replica of $p=1000$ independent measurements
and the true value of the phase shift $\theta$. The vertical bars
are the mean square fluctuations $\sigma_{est}^2=\langle
(\Theta_{est}-\langle \Theta_\text{est} \rangle)^2\rangle$. The
result $\sigma_{est}\gg(\theta-\Theta_{est})$ proves that our
protocol provides an unbiased experimental phase
estimation.}}\label{estimatore}
\end{figure}

The first part of the experiment consists of the calibration of
the interferometer. At different, known, values of the phase shift,
we measured $N_c$ and $N_d$ for each of $200,000$ laser pulses.
This procedure allows us to determine experimentally both
$P(N_c,N_d|\phi)$ and $P(\phi | N_c,N_d)$. In Fig.(\ref{R}) we
compare the ideal and the experimental phase distributions. The
agreement is quite good, and the discrepancies can be attributed
to imperfect photon-number discrimination by the detectors.
To fit the data in Fig.(\ref{R}), we introduce the probability
$P(N_c, N_d| N_c', N_d')$ to measure $N_c$ and $N_d$
when $N_c'$ and $N_d'$ photons were really present.
Taking this into account, the experimental phase probability distribution is
\be \label{fitP}
P_\text{fit}(\phi|N_c, N_d) = \sum_{N_c', N_d'} P(\phi|N_c', N_d')  P(N_c', N_d'| N_c, N_d),
\ee
where $P(\phi|N_c', N_d')$ are the ideal probabilities Eq.(\ref{coherent_dist}).
The weights $P(N_c', N_d'| N_c, N_d)$ can be retrieved from a fit of
the experimental calibration distributions $P(\phi|N_c,N_d)$, see Fig.(\ref{R}).
The quantity $P(N_c, N_d| N_c, N_d)$, equal to one in the ideal
case, is 0.54 in Fig.(\ref{R}A) (corresponding to the worst case
among all distributions), 0.67 in (\ref{R}B) and (\ref{R}C), and
0.87 in (\ref{R}D).

\begin{figure}
\begin{center}
\includegraphics[scale=0.34]{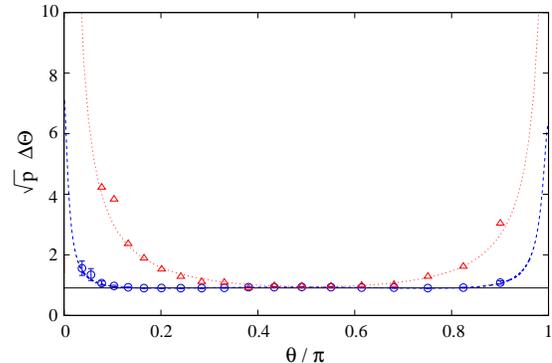}
\end{center}
\caption{\small{(color online). Phase
sensitivity as a function of the true value of the phase shift.
Circles are obtained from Bayesian distributions Eq.(\ref{Pp}), with $p=1000$.
The error bars give the fluctuations of $\sqrt{p} \, \Delta \Theta$ obtained
with 150 independent replica of the experiment.
The solid black line is the theoretical prediction Eq.(\ref{shot_noise}).
The dashed blue line is the CRLB calculated with the
experimental distributions. Triangles are the
uncertainty obtained with a generalization of the estimator
Eq.(\ref{estimator}) taking into account the experimental
imperfections, while the dotted red line is the phase
sensitivity predicted by Eq.(\ref{momentum_formula}).}}\label{errore}
\end{figure}

After the calibration, we can proceed with the Bayesian phase
estimation experiment. For a certain value of the phase shift, we
input one laser pulse and detect the number of photons $N_c$ and
$N_d$. We repeat this procedure $p$ times obtaining a sequence of
independent results $\{N_c^{(i)}, N_d^{(i)} \}_{i=1\ldots p}$. The
$p$ photon-number measurements comprise a single phase estimation.
The overall phase probability is given by the product of the
distributions associated with each experimental result:
\be\label{Pp}
P\big(\phi|\{N_c^{(i)}, N_d^{(i)} \}_{i=1\ldots
p}\big) \propto \prod_{i=1}^{p} P_\text{fit}\big(\phi| N_c^{(i)},
N_d^{(i)} \big).
\ee
The phase estimator $\Theta_\text{est}$ is given
by the mean value of the distribution,
\mbox{$\Theta_\text{est}=\int_{0}^{\pi}\ud \phi \, \phi \,
P\big(\phi|\{N_1^{(i)}, N_2^{(i)} \}_{i=1 \ldots p}\big)$}, and
the phase uncertainty $\Delta \Theta$ is the $68.27\%$ confidence
interval around $\Theta_\text{est}$. An example of
$P\big(\phi |\{N_c^{(i)}, N_d^{(i)} \}_{i=1 \ldots p}\big)$ is
given in Fig.(\ref{p}), for $\theta/\pi=0.24$ and for different
values of $p$. Since the average number of photons of the coherent
input state is small, for $p \sim 1$ the phase uncertainty is of
the order of the prior knowledge, $\Delta \Theta \simeq \pi$. As
$p$ increases, the probability distribution becomes Gaussian and
the sensitivity scales as $\Delta \Theta \propto 1/\sqrt{p}$, in
agreement with the central limit theorem.

In Fig.(\ref{estimatore}), we show the difference between the mean value of the phase
estimator $\langle \Theta_\text{est} \rangle$, obtained from 150 phase
estimations, each with $p=1000$, and the true value of the phase shift. 
The bars are the mean square fluctuation. 
The important result is that our protocol provides an experimentally unbiased
phase estimation over the entire phase interval.

The main result of this Letter is presented in Fig.(\ref{errore}).
We show the phase sensitivity for different values of the phase
shift $\theta$, calculated from the distribution Eq.(\ref{Pp})
with $p=1000$ photon-number measurements. The circles are the mean
value of $\Delta \Theta$, and the bars give the corresponding mean
square fluctuation, obtained from 150 independent phase
measurements. The dashed blue line is the CRLB calculated with the
experimental probability distributions, $\Delta
\theta_{\mathrm{fit}}=1/\sqrt{p F_{\mathrm{fit}}(\theta)}$, where
$F_{\mathrm{fit}}(\theta)=\sum_{N_1,N_2}\frac{1}{P_{\mathrm{fit}}(N_1,
N_2|\theta)} (\frac{\partial
P_{\mathrm{fit}}(N_1,N_2|\theta)}{\partial\theta})^2$
\cite{nota3}. For $0.1\lesssim \theta/\pi \lesssim 0.9$, it
follows the theoretical prediction (solid black line),
Eq.(\ref{shot_noise}), where $\bar{n}=1.08$ has been
independently calculated from the collected data. Around
$\theta=0,\pi$, where the photons have higher probability to exit
through the same output port, $\Delta \theta_{\mathrm{fit}}$
increases as a consequence of the decreased sensitivity of 
our detectors to higher photon number states.
Even though the phase sensitivity of our apparatus becomes worse near $\theta=0,\pi$,
it never diverges. 
Triangles show the phase uncertainty obtained
with Eq.(\ref{estimator}), but taking into account the
experimental noise. The estimator is obtained by inverting the
equation $M_p=\bar{n}\cos(a+\theta)+b$ \cite{nota1}. This strategy
provides an unbiased estimation with a sensitivity close to the
one predicted by Eq.(\ref{momentum_formula}) (dotted red line). The
reason for the superior performance of the Bayesian protocol can
be understood by noticing that, in Eq.(\ref{momentum_formula}),
the phase estimate is retrieved only from the measurement of the
photon number difference, which, simply does not exploit all of
the available information.

In \cite{Yurke_1986} YMK first proposed a generalization of the
estimator Eq.(\ref{estimator}) to take into account the whole
information in the output measurements. Their estimator,
$\Theta_\text{YMK} = \arccos[(N_c-N_d)/(N_c+N_d)]$, gives a phase
independent sensitivity, $\Delta \Theta = 1/\sqrt{N_c+N_d}$.
Notice that $\Theta_\text{YMK}$ coincides with the Maximum
Likelihood estimator in the ideal, noiseless, MZ interferometer.
However, it is not obvious how to generalize $\Theta_\text{YMK}$
for real interferometry, where classical noise is present and the
YMK estimator is different from $\Theta_{ML}$. In general, because
of correlations between $N_c-N_d$ and $N_c+N_d$, this estimator
becomes strongly \emph{biased} in the presence of noise as we have
verified using our experimental data \cite{Pezze}. Moreover, the
YMK estimator cannot be extended when both input
ports of the interferometer are illuminated. Conversely, the
Bayesian analysis holds for general inputs and, in particular, it
predicts a phase-independent sensitivity when squeezed vacuum is
injected in the unused port of the MZ \cite{Pezze}, which reaches
a sub shot-noise sensitivity \cite{Caves_1981}.  It should be
noted that detection losses again become important when attempting
to use nonclassical light to overcome the shot-noise
limit Eq.(\ref{shot_noise}).

In \cite{Hradil_1996}, Hradil et al. used a Bayesian approach for a
Michelson-Morley neutron interferometer (single output detection)
and discussed theoretically the MZ. Their analysis was based on
specific assumptions about the interferometric classical noise
which are not satisfied in the case
discussed in this Letter. Different approaches with adaptive
measurements \cite{Berry_2000} and positive operator value
measurements \cite{Sanders_1995} have been also suggested. While
these strategies might be important for interferometry at the
Heisenberg limit, they are not necessary in our case.

In conclusion, we have presented a Bayesian phase estimation
protocol for a MZ interferometer fed by a single coherent state.
The protocol is unbiased and provides a phase sensitivity that
saturates the ultimate Cramer-Rao uncertainty bound imposed by
quantum fluctuations. We have been able to implement the protocol
with two photon-number-resolving detectors at the output ports of
a weakly-illuminated interferometer. Yet, the method can be
generalized to the case of high intensity laser interferometry and
photodiode detectors. In this case, the limit
Eq.(\ref{shot_noise}) becomes harder to achieve because of larger
electronic noise and lower photon number resolution, however it
should still be possible to demonstrate a phase independent
sensitivity. 
Our results are of importance to quantum
inference theory and show that the MZ interferometer does not
require phase-locking in order to reach an optimal sensitivity.

Acknowledgment. This work has been partially supported by the US
DoE and NSF Grant No. 0304678.


\begin{thebibliography}{50}
\bibitem{nota2}
The Quantum Mechanical probability to detect $N_c$ and $N_d$ particles
at the output port $c$ and $d$, respectively, is given by
$P(N_c, N_d|\phi) = | (\langle N_c|\langle N_d|) e^{-i \phi \hat{J}_y} | \psi_{inp} \rangle |^2$,
where $\hat{J}_y=(\hat{a}^{\dag}\hat{b}-\hat{b}^{\dag}\hat{a})/2i$,
with $\hat{a},\hat{b}$ bosonic annihilation operators.
For the configuration studied in this Letter
$\label{Pout}
P(N_c, N_d|\phi) = \frac{|\alpha|^{2N} e^{-|\alpha|^2}}{(N_c)!(N_d)!}
\big(\sin \frac{\phi}{2}\big)^{2N_c} \big(\cos \frac{\phi}{2}\big)^{2N_d}$.
\bibitem{nota3} The distributions
$P_{\mathrm{fit}}(N_1,N_2|\theta)$
are obtained from Eq.(\ref{fitP}) with the Bayes theorem.
\end{thebibliography}
\end{document}